\documentclass[journal, 12pt]{IEEEtran}
\usepackage{amsmath,amsfonts}
\usepackage{algorithmic}
\usepackage{array}
\usepackage[caption=false,font=normalsize,labelfont=sf,textfont=sf]{subfig}
\usepackage{textcomp}
\usepackage{stfloats}
\usepackage{url}
\usepackage{verbatim}
\usepackage{graphicx}
\usepackage{color}
\def\BibTeX{{\rm B\kern-.05em{\sc i\kern-.025em b}\kern-.08em
    T\kern-.1667em\lower.7ex\hbox{E}\kern-.125emX}}
\usepackage{balance}

\newcommand{\reff}[1]{Fig.~\ref{#1}}

\begin{document}
\title{Blockchain-Enabled Management Framework for Federated Coalition Networks}
\author{Jorge Álvaro González, Ana María Saiz García, and Victor Monzon Baeza,~\IEEEmembership{Senior Member,~IEEE}  
\thanks{J. Alvaro Gonzalez, Ana María Saiz Garcia and V. monzon Baeza are with University Oberta Catalunya, Barcelona, Spain. \textit{Corresponding author: Jorge Álvaro González (jalvarogo@uoc.edu).}
}}

\markboth{Journal of \LaTeX\ Class Files,~Vol.~X, No.~X, September~2024}%
{How to Use the IEEEtran \LaTeX \ Templates}

\maketitle


\begin{abstract}
In a globalized and interconnected world, interoperability has become a key concept for advancing tactical scenarios. Federated Coalition Networks (FCN) enable cooperation between entities from multiple nations while allowing each to maintain control over their systems. However, this interoperability necessitates the sharing of increasing amounts of information between different tactical assets, raising the need for higher security measures. Emerging technologies like blockchain drive a revolution in secure communications, paving the way for new tactical scenarios. In this work, we propose a blockchain-based framework to enhance the resilience and security of the management of these networks. We offer a guide to FCN design to help a broad audience understand the military networks in international missions by a use case and key functions applied to a proposed architecture. We evaluate its effectiveness and performance in information encryption to validate this framework.
\end{abstract}


\section{Introduction}

The digital era and advancing technology are revolutionizing both civilian and military environments. This technological shift is reshaping how industries operate and introducing new ways to enhance efficiency, communication, and control across various sectors, especially in defense \cite{Victor1_New, mot1}. Security has always been the top priority in military missions and tactical scenarios \cite{10017647}. Today, security is viewed from two critical perspectives: the protection of human lives and the security of information. As our world becomes more interconnected and digitized, sensitive information increases exponentially. This growing flow of data heightens the risks of cyber threats, intrusions, theft, and other dangers that, if unaddressed, can ultimately compromise the safety of individuals as well.

New technologies are being integrated into the military landscape to meet these rising challenges \cite{telecomLaura},\cite{joseNEt}. These technologies are crucial for addressing the ever-evolving demands of modern tactical environments and are beneficial for dynamic environments such as battlefields. Their role is to improve security measures, safeguard sensitive information, and protect critical assets and personnel in increasingly complex operational scenarios. Among these technologies, we highlight software-defined networks, enabling flexibility and scalability, adapting many situations as needed \cite{9486399}. Others, like blockchain, may be a powerful tool for enhancing security. By its decentralized and tamper-resistant nature, blockchain provides a robust solution for securing sensitive data, ensuring integrity, and enabling transparent transactions across networks. Its potential to revolutionize military operations \cite{10625605} by ensuring secure communication channels and resilient data sharing makes it an ideal candidate for adoption in modern tactical environments.

Future missions increasingly require joint actions, making interoperability a key concept, which refers to the ability of different national systems and networks to work together efficiently without compromising security or confidentiality \cite{9282382}. Organizations like the North Atlantic Treaty Organization (NATO) play a pivotal role in the success of cooperation and interoperability for international missions where coalition among members is a duty. The need for seamless communication and coordination among different countries has become critical in coalition-based operations. FCNs were developed to address this challenge \cite{10253524}. FCNs are communication infrastructures that enable interoperability and cooperation among multiple entities or nations in joint missions, particularly in military or defense contexts. These networks allow different armed forces, security agencies, or allied organizations to securely share information and resources while maintaining independence and control over their systems. FCNs are also flexible and capable of scaling and adapting to mission requirements. They provide a robust infrastructure that can integrate with emerging technologies like blockchain to enhance security, traceability, and data management. This integration improves the protection of critical information and builds trust among the involved parties. The management of FCNs would be carried out from Command and Control (C2) centers. However, despite the advancements in FCNs, there remains a gap in integrating blockchain technology into these networks to bolster security and resilience further from the point of view of the management in C2. Current research does not fully explore how blockchain can enhance the management and protection of FCN. 
 
In this context, we analyze the challenges and benefits of blockchain in FCN. We propose a blockchain-based framework designed to fill this gap in this work, offering a secure, efficient FCN management solution that strengthens network resilience and information security.  
  
The rest of the paper is organized as follows: Section 2 reviews the main FCNs. Section 3 explores the benefits of blockchain in FCNs. Section 4 presents the application scenario. Section 5 details the functions incorporated in our proposal. Section 6 describes the developed architecture. Finally, Section 7 analyzes the results, and Section 8 provides the conclusions.

\section{Federated Networks Overview}

Resource federation has gained popularity as it creates a federated system with enhanced aggregate capability, where the overall capacity exceeds the sum of individual resources. This concept extends to network federation, or Federated Networks (FEDNETs) \cite{niemegeers2005fednets}, which are ad-hoc and temporary networks collaborating under a unified framework of rules, services, and policies to achieve common objectives.

\subsection{Afghan Mission Network (AMN)}
The first mission-centric application of the FEDNET concept was the \textbf{Afghan Mission Network (AMN)}. It resulted from years of collaboration among Troop Contributing Nations (TCNs) who deployed infrastructure, services, and troops during counter-terrorist operations in Afghanistan from 2001 to 2021.
During these operations, the TCNs were forced to collaborate in a technical heterogeneous environment, and in 2012, NATO Military Committee(NATO MC) approved the request to establish a common domain where the TCNs could share services and information. This common domain could only be achieved by domain and network federation\cite{NATOC2COE_2012}. 

AMN was also conceived as an extended C2 system, as it would manage information exchange and pose as a framework for future definitions of systems. Consequently, the main objectives that future systems should achieve are:
\begin{itemize}
    \item Information sharing: it is critical for the mission to achieve an environment in which the TCNs are able to send and share information between military and civil organizations.
    \item Management and governance: the involved stakeholders should use a commonly agreed set of policies, protocols, technologies, and standards as an interoperability framework during deployment.
\end{itemize}

\subsection{Federated Mission Networking (FMN)}

The AMN's evolution led to the FMN initiative, initially called Future Mission Networking and later Federated Mission Networking. This concept aims to enhance interoperability and readiness among the various forces and entities involved in missions \cite{pullen2020nato}. FMN is still an ongoing effort which aims to:
\begin{itemize}
    \item Improve the strategic advantage by optimizing the deployment requirements for the NATO verification and validation activities.
    \item Continuous improvement and adaptation to the technical challenges transforming the mission landscape. This way, the NATO forces will stay in the technological avant-garde.
\end{itemize}
FMN efforts are organized into "spirals," each targeting specific capabilities and technologies, such as mobile communications, to be implemented by the end of each development phase. The initiative aims to define a coordinated set of processes, organizations, training, technologies, and standards for NATO nations and allies. Each component must enable each Spiral's required services, sub-services, and capabilities to achieve the desired interoperability.

\section{Benefits of Blockchain in FCN}
FCNs often operate in dynamic and complex environments where traditional centralized management systems may lack the flexibility and security required. With its decentralized architecture, Blockchain addresses these issues by eliminating the need for a central authority, thereby reducing the risk of a single point of failure. Originally developed for cryptocurrencies like Bitcoin, blockchain has evolved into a versatile tool for enhancing security, transparency, and efficiency in various applications. Its decentralized nature and immutable and transparent transaction records make it an ideal candidate for managing FCNs from C2 systems.

A blockchain-enabled management framework offers several key advantages for FCNs:
\begin{enumerate}
    \item Enhanced Security: Blockchain provides a tamper-proof ledger for all transactions and interactions within the network. Each transaction is cryptographically secured and linked to the previous one, creating a difficult-to-alter chain of blocks. This ensures data integrity and protects against unauthorized access and tampering.

    \item Increased Transparency and Trust: Transactions are recorded on a public or permissioned ledger accessible to all authorized participants. This transparency allows all actions within the network to be visible and verifiable, fostering trust and collaboration among coalition members. This capability is particularly valuable in defense and military operations where trust is critical.

    \item Automation with Smart Contracts: Blockchain technology enables the use of smart contracts—self-executing agreements with terms written directly into code. Smart contracts can automatically enforce rules, execute transactions, and manage access control based on predefined conditions. This reduces administrative overhead, minimizes human errors, and ensures compliance with established protocols.
\end{enumerate}

Despite these advantages, integrating blockchain into FCNs presents challenges. Issues such as scalability, interoperability, and the need for consensus mechanisms tailored to coalition networks must be addressed. Additionally, the computational and energy requirements for maintaining a blockchain can be significant, necessitating the development of efficient protocols and mechanisms for sustainable operation.

Blockchain technology has the potential to revolutionize the management of federated coalition networks. By providing a secure, transparent, and efficient management framework, blockchain can enhance collaboration, trust, and operational efficiency among diverse entities. This paper explores the practical application of blockchain in FCNs, analyzes its benefits, addresses associated challenges, and proposes solutions for successful integration. As the digital landscape evolves, adopting innovative technologies like blockchain will be crucial for ensuring the robustness and resilience of FCNs. 

\section{Scenario and Use Case description}
\begin{figure}[h]
    \centering
    \captionsetup{justification=centering}
    \includegraphics[width=1\columnwidth]{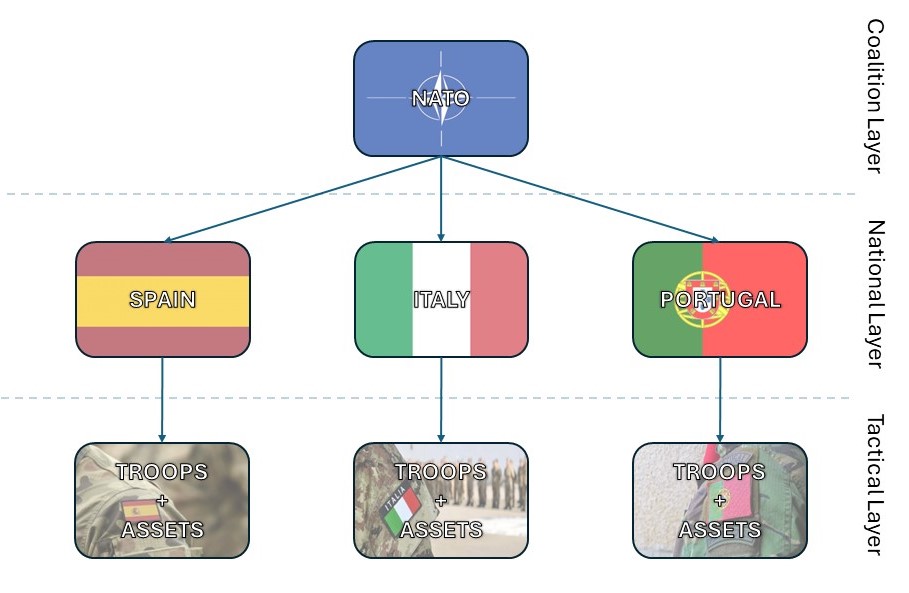}
    \caption{Coalition hierarchy.}
    \label{fig:fig1}
\end{figure}

The scenario considered for our proposal involves forces from three allied countries, one adversarial force, and a neutral territory where all these forces, including neutral parties, are present.
Focusing on the coalition deploying FMN-ready technologies, the forces will be deployed following a three-tier hierarchy shown in \reff{fig:fig1}.
\begin{enumerate}
    \item \textbf{Tier one, or Coalition Layer}: at the top of Fig. \ref{fig:fig1}.  It includes a supranational entity centralizing coalition C2 operations. This layer represents a Joint Force Command (JFC) for this illustrative example. 
    \item \textbf{Tier two, or National Layer}: in the central section of \reff{fig:fig1}. It involves the national C2 systems that will transpose the coalition orders to the national components and forces. Here, the national layer involves three allied nations: Spain, Italy, and Portugal.
    \item \textbf{Tier three, or Tactical Layer}: at the bottom of \reff{fig:fig1}. It executes the mission orders and includes the national troops and assets. Each component of this layer is federated with its own national C2.
\end{enumerate}
The hierarchy operates across different functional levels: the Coalition layer sets high-level commands, the national layer translates them into executable actions, and the tactical layer carries out these commands (e.g. mission orders or asset configurations).

\begin{figure}[h]
    \centering
    \captionsetup{justification=centering}
    \includegraphics[width=0.9\columnwidth]{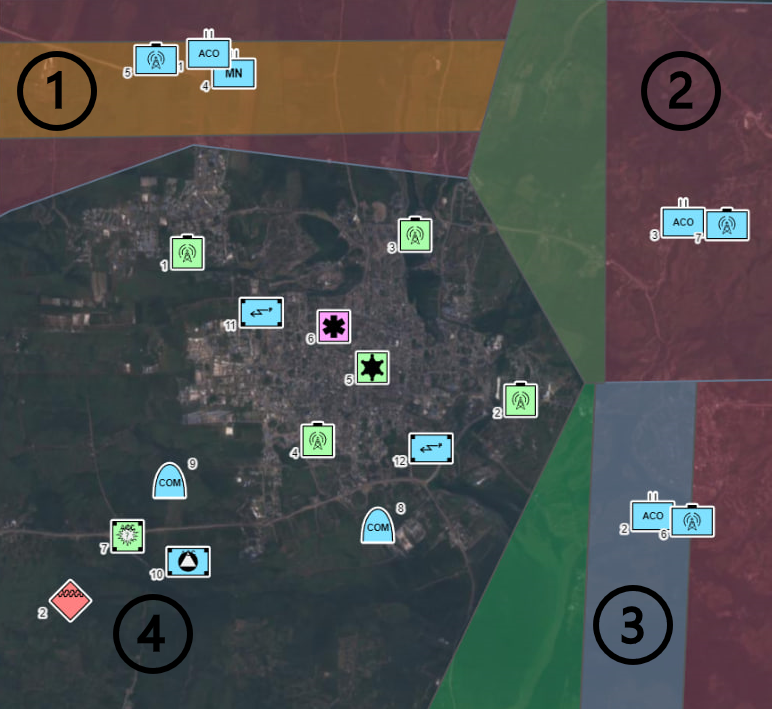}
    \caption{Scenario location.}
    \label{fig:fig2}
\end{figure}

In the fictional city scenario, Spanish troops are stationed in the north, Portuguese forces in the northeast, and Italian forces in the southeast, as shown in \ref{fig:fig2} (numbers 1, 2, and 3). All allied forces are engaged in routine peacekeeping tasks within the city.

The use case describes an incident that eventually occurs in the southwest, labeled with 4 in \reff{fig:fig2}. The allied forces come to secure the area, and as this zone is considered a non-secure area, communications will be carried out through a swarm of drones rather than through the infrastructure present in the area in order to ensure the transmitted information. During the operations, one of the communication drones begins to malfunction and crashes in a distant area under the control of enemy forces.

The proposed framework highlights areas 1 to 3 for this use case (see Figure \ref{fig:fig2}) as key zones that would benefit from implementing Blockchain in the FMN environment as follows:
\begin{itemize}
    \item Before the operation, when forces are deployed or a new ally joins, the IT infrastructure must be configured. Blockchain can involve all forces and authorized stakeholders to ensure proper device configuration.
    \item During the operation, blockchain can secure communications between the troops that develop the mission, including tactical exchanges, order distribution, mission reports, and device log management. 
    \item After the operation, the system would ensure the access, availability, and integrity of mission reports for their further analysis.
\end{itemize}

\section{Functions Requirements Description}

This section outlines the key functions necessary to support the blockchain-powered management system in FCN according to the previous scenario. The proposed structure enhances security, transparency, and efficiency using blockchain technology, focusing on seamless communication, coordination, and resource management across all coalition levels, from strategic to tactical units.

We categorize the functionalities into two main types: \textbf{Service Policy-} and \textbf{Network-based functionalities}. These categories are crucial for ensuring secure operations, optimizing resource allocation, and enhancing interoperability in federated networks, thereby supporting successful mission outcomes.

\subsection{Service Policy Functionalities}

Enforcing security policies on utilizing blockchain is crucial for protecting valuable assets and ensuring operational integrity in a multi-nation military conflict zone. Blockchain enhances this process by providing a decentralized, transparent, and immutable method for defining, applying, and auditing security policies. This technology streamlines establishing security protocols, their implementation on network devices, and the verification of adherence, offering a secure and distributed platform for efficient policy management.

\subsubsection{Minimum Necessary Conditions} 
\begin{itemize}
    \item Using blockchain enables the implementation of key security management tools, such as configuration management systems, Intrusion Detection Systems (IDS), and access control mechanisms, which are vital for enforcing security in dynamic military environments.
    \item Blockchain’s immutable record ensures that all policy changes and compliance checks are securely documented and auditable, allowing real-time monitoring and verification of adherence.
\end{itemize}

\subsubsection{Key assets} 
\begin{itemize}
    \item Safeguarding Critical Assets and Data: Security measures are vital for protecting information integrity and confidentiality in conflict zones. Blockchain technology supports these measures by ensuring adherence to security standards across a decentralized network. 
    \item Flexibility and Scalability: Blockchain enables large-scale adaptation of security measures, which is crucial for managing complex multinational military operations. Its distributed structure ensures continuous policy enforcement without relying on a central authority, which is vital for high-risk missions. 
    \item Ingrained Audit and Conformity: Blockchain’s auditing features ensure continuous monitoring and validation of policy adherence, quickly identifying and resolving discrepancies. This reduces security breaches and enhances network resilience. 
\end{itemize}
Using blockchain to enforce policies in military services during operations enhances security and operational integrity. Its decentralized and transparent nature ensures consistent rule enforcement and rapid detection of unauthorized access or policy breaches.

\vspace{-0.1cm}
\subsection{Network Functionalities}

Communication and data integrity are crucial in conflict zones, where multiple countries shape the network. Blockchain enhances Public Key Infrastructure (PKI) management by decentralizing and securing the process. Instead of relying on a central authority, blockchain distributes trust across the network, with each node (element or component of FCN) maintaining and verifying public keys. This approach strengthens security by eliminating single points of failure and making it harder for malicious actors to compromise the system. In this case, tactical networks and operations are seen as empowered.

\subsubsection{Minimum Necessary Conditions}  
\begin{itemize}
    \item Implementing blockchain-based PKI requires strong encryption and secure key management to guard against cyber threats. Access controls at all network nodes ensure that only authorized entities can manage certificates, maintaining system integrity in critical military environments.
    \item The PKI system must handle many devices, users, and transactions typical in military deployments. Blockchain’s distributed architecture supports this scalability, enabling the system to process high certificate issuance and validation requests without compromising performance.
    \item Blockchain’s decentralized structure provides inherent redundancy, ensuring the system remains operational even if certain nodes fail. This resilience is vital in conflict zones, where network reliability may be compromised due to operational disruptions or attacks.
    \item To meet the demands of military networks, the blockchain-based PKI must perform certificate-related operations such as validation and issuance with minimal latency. High-performance capabilities are crucial for ensuring real-time communication and timely decision-making in tactical situations.
\end{itemize}

\subsubsection{Key assets} 
\begin{itemize}
    \item Distributed Trust: Blockchain removes the need for a central authority by spreading trust across all network nodes. This reduces vulnerabilities and is especially useful in collaborative military operations, where no single nation should control the entire PKI system.
    \item Immutability and Auditability: Blockchain's immutable ledger ensures that public keys and certificates cannot be altered once recorded. This creates a secure and auditable history of all PKI transactions, allowing real-time tracking of issuance, renewal, and revocation, which is crucial for maintaining integrity in sensitive military environments.
    \item Automated Certificate Management: smart contracts automate tasks like certificate issuance, renewal, and revocation, reducing manual work and enhancing efficiency and security. This automation also promptly revokes certificates, minimizing unauthorized access and boosting overall network security. 
\end{itemize}

\section{Proposed FCN Architecture}
This proposal is based on the NATO Architecture Framework v4 (NAFv4) guideline \cite{NAFref}, which is ideal for defining architectures in both business and military contexts. We selected the following viewpoints (C1, S1) from NAFv4 for our system to ensure the appropriate level of detail for our proposal:
\begin{itemize}
    \item \textbf{C1 - Capability Taxonomy}: this viewpoint is concerned with identifying the capabilities and their organization, independent of their implementation.
    In this case, this viewpoint has been used to identify and organize the internal capabilities, meaning only those capabilities related to the proposed Blockchain-based system. Following the hierarchy and structure presented in \reff{fig:fig1}, the capabilities identified can be linked to a certain tier of the federation. The capability related to the data analysis and the lessons learned is transversal to all the mission planning and organization tiers, as shown in \reff{fig:fig3}. 
\begin{figure}[h]
    \centering
    \captionsetup{justification=centering}
    \includegraphics[width=1\columnwidth]{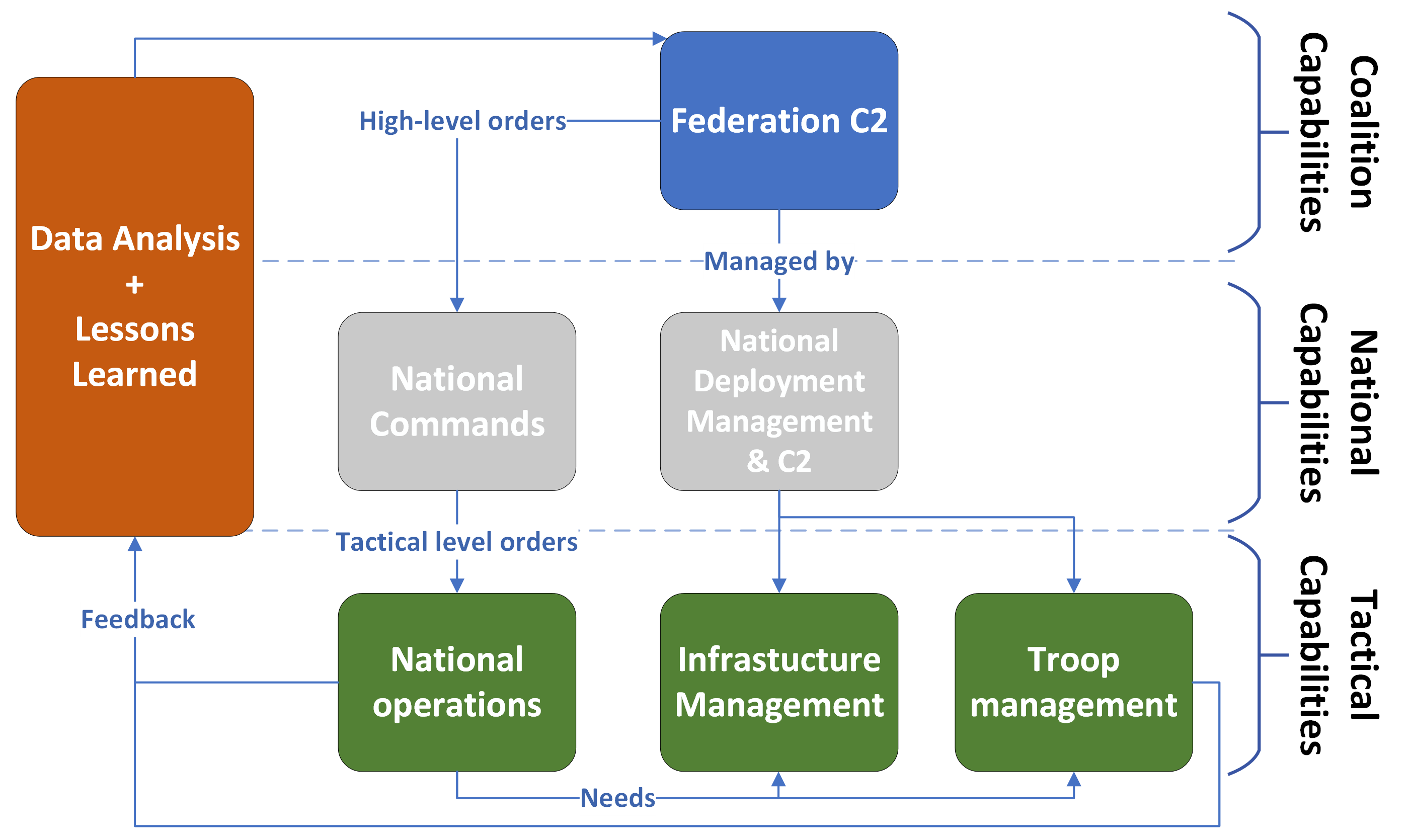}
    \caption{C1 viewpoint.}
    \label{fig:fig3}
\end{figure}
    \item \textbf{S1 - Service Taxonomy}: this viewpoint is concerned with identifying the service specifications and their organization. In our case, the viewpoint shown in \reff{fig:fig4} has been used to define the services that would be required to enable the particular capability of Infrastructure Management identified in the previous viewpoint C1. 
    \begin{figure}[h]
    \centering
    \captionsetup{justification=centering}
    \includegraphics[width=1\columnwidth]{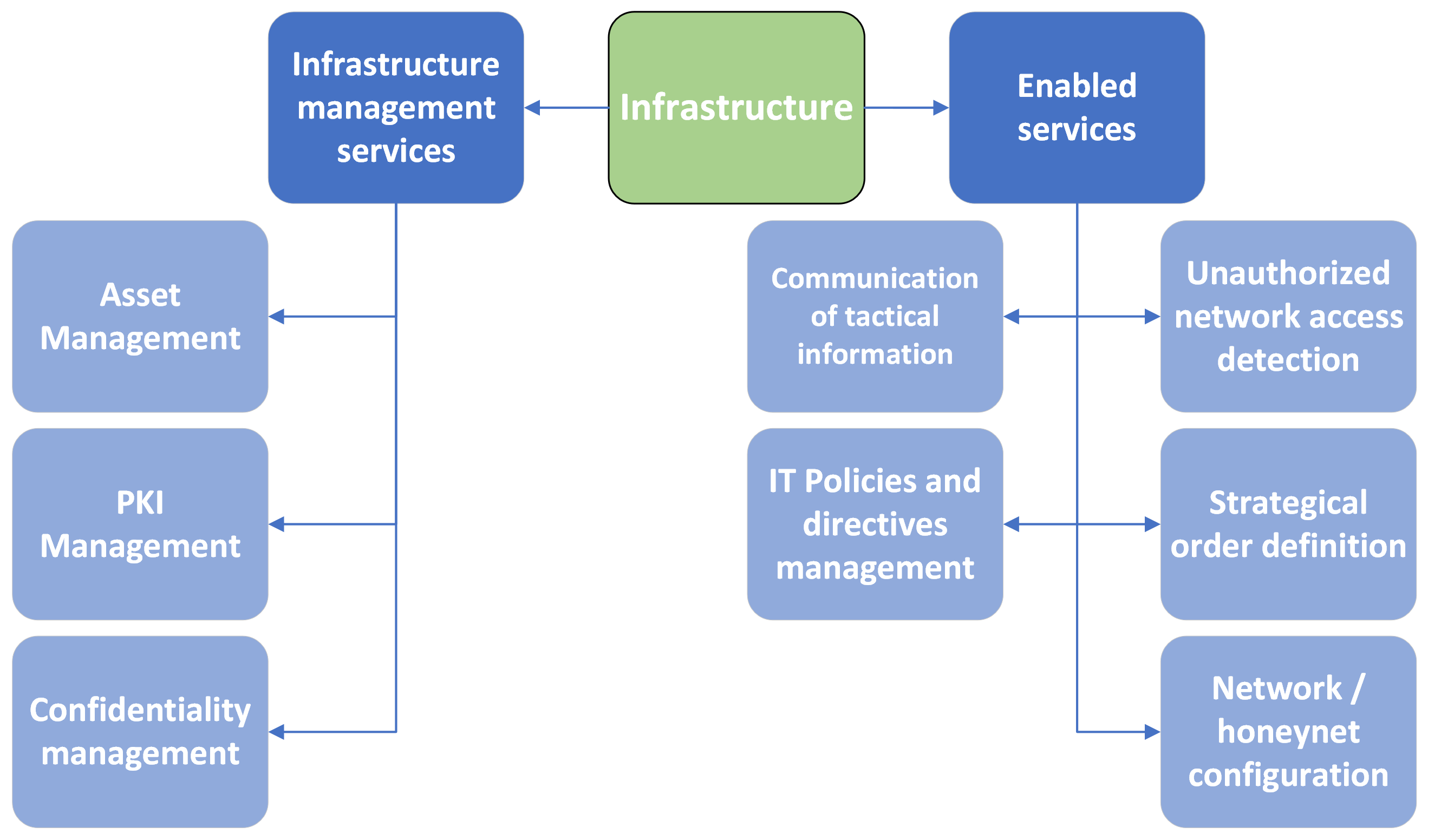}
    \caption{S1 viewpoint, Infrastructure management capability.}
    \label{fig:fig4}
    \end{figure}
    As this viewpoint shows, IT infrastructure management includes the management of public key infrastructure, the network and assets configuration, and other management and enabled services.
\end{itemize}
The identified services are associated with different phases of the use case exposed in the scenario according to (1) During the deployment of troops, the networks and assets should be properly configured; (2) the deployed assets must be assigned to a user-authorized in the system, and their keys properly enrolled in the PKI, and (3) the troops should receive the mission order ensuring their confidentiality. 
 
\section{FCN Proof-of-concept Evaluation}

Various alternatives were considered to evaluate the proposal, and the final proof of concept (PoC) was developed on the Avalanche Testnet. Avalanche provides an excellent environment for deploying smart contracts. As requirements for functionalities in Section V, the deployed contracts focus on Policy Enforcement and PKI Management functionalities. The PoC includes the following functions:
\begin{itemize}
    \item \textbf{Policy enforcement}
    \begin{enumerate}
        \item Creation and initialization of policy sets; system administrators can call this function.
        \item Retrieve policy status and configuration; this function can be called by any authorized user.
        \item Request policy modification; this function can be called by any authorized user.
        \item Modify policy or directive; this function can be called by system administrators.
        \item Evaluate and apply or deny modification requests; this function can be called by system administrators.
    \end{enumerate}
    \item \textbf{PKI management}
    \begin{enumerate}
        \item Publish public key; this function can be used by any authorized user.
        \item Update published public key; this function can be used by any authorized user.
        \item Retrieve published public key; any authorized user can use this function.
        \item Revoke published public key; this function can be used by system administrators.
        \item Authorize published public key; this function can be used by system administrators.
    \end{enumerate}
\end{itemize}

After deploying on Avalanche\footnote{Note that full implementation would require developing additional functionalities depending on the country to provide comprehensive services.}, the results of the contracts were evaluated. All functions were tested with authorized, unauthorized, and administrator accounts. \reff{fig:fig4} illustrates how transactions are recorded on the blockchain. The figure highlights several fields in blue, such as the function called transaction details and user identifiers, all securely encrypted. In contrast, highlighted in green, authorized users could access the requested information.

\begin{figure}[h]
    \centering
    \captionsetup{justification=centering}
    \includegraphics[width=1\columnwidth]{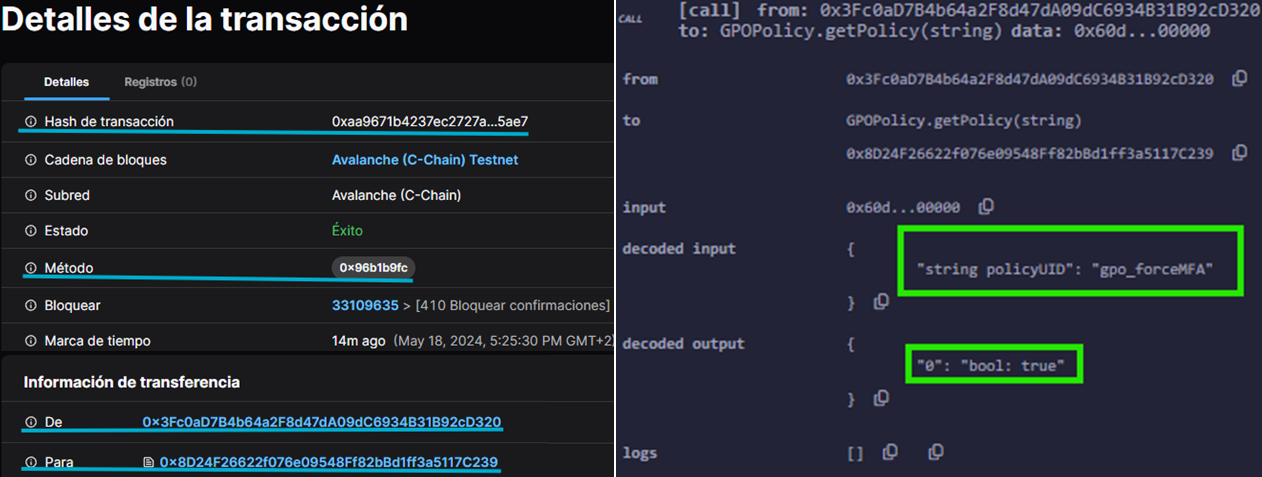}
    \caption{Transaction details validation.}
    \label{fig:fig4}
\end{figure}


The tests demonstrate that the functions implemented through smart contracts operate as intended, confirming the system's viability. Furthermore, the fundamental security dimensions are ensured, with information properly encrypted against unauthorized access. Thanks to the nature of blockchain and the validating nodes, the system's availability and integrity are guaranteed.

\section{Conclusion}

This proposal highlights significant interoperability opportunities for enhancing FCN deployments and provides a guide to understanding FCNs, including the benefits and challenges of using blockchain. Through an exemplary scenario, it demonstrates how blockchain can improve security, transparency, and efficiency in multi-national military operations.The proof-of-concepts and tests confirm the potential of the proposed FCN system.

\balance
\bibliographystyle{IEEEtran}
\bibliography{References}


\begin{IEEEbiographynophoto}{Jorge Alvaro Gonzalez} 
earned his M.Sc. in the Open University of Catalonia (2024), respectively, and an M.Sc. in ICT Cybersecurity from the European University of Madrid (2022). Previously, he worked in the Defence industry and is now a cybersecurity engineer at Alstom in the OT railway industry.
\end{IEEEbiographynophoto}
\vskip -2\baselineskip plus -1fil
\begin{IEEEbiographynophoto}{Ana Saiz García} 
received her M.Sc. from the Open University of Catalonia (2024). She is currently employed in the Defence sector at Indra, where she is involved in a range of projects for both NATO and the Spanish Ministry of Defence.
\end{IEEEbiographynophoto}
\vskip -2\baselineskip plus -1fil
\begin{IEEEbiographynophoto}{Victor Monzon Baeza} 
received his Ph.D. degree in Electrical Engineering from the University Carlos III of Madrid, Spain, in 2019. He was a software developer in the Defense sector and is currently working in Sateliot, Spain, as a Payload Architect and Teacher in UOC.
\end{IEEEbiographynophoto}

\end{document}